\documentclass[aps,prl,twocolumn,superscriptaddress,showpacs]{revtex4-1}
\usepackage{amsmath,amssymb}
\usepackage{graphicx}
\usepackage{bm}
\usepackage{mathrsfs}
\usepackage{textcomp}
\usepackage[normalem]{ulem}

\begin{document}

\title{A Theory of Shape-Shifting Droplets}
\author{Pierre A. Haas}
\author{Raymond E. Goldstein}
\email{R.E.Goldstein@damtp.cam.ac.uk}
\affiliation{Department of Applied Mathematics and Theoretical Physics, Centre for Mathematical Sciences, \\ University of Cambridge, 
Wilberforce Road, Cambridge CB3 0WA, United Kingdom}
\author{Stoyan K. Smoukov}
\email{sks46@cam.ac.uk}
\affiliation{Department of Materials Science and Metallurgy, University of Cambridge, 27 Charles Babbage Road, Cambridge CB3 0FS, United Kingdom}
\author{Diana Cholakova}
\author{Nikolai Denkov}
\affiliation{Department of Chemical and Pharmaceutical Engineering, Faculty of Chemistry and Pharmacy, \\ University of Sofia, 1164 Sofia, Bulgaria}
\date{\today}%
\begin{abstract}
Recent studies of cooled oil emulsion droplets uncovered transformations into a host of flattened shapes with straight edges and sharp corners, 
driven by a partial phase transition of the bulk liquid phase. Here, we explore theoretically the simplest geometric competition between this 
phase transition and surface tension in planar polygons, and recover the observed sequence of shapes and their statistics in qualitative 
agreement with experiments. Extending the model to capture some of the three-dimensional structure of the droplets, we analyze the 
evolution of protrusions sprouting from the vertices of the platelets and the topological transition of a puncturing planar polygon.

\end{abstract}

\pacs{82.70.-y, 81.30.-t, 05.70.Ln, 64.75.-g}

\maketitle

\begin{figure}[b]
\includegraphics{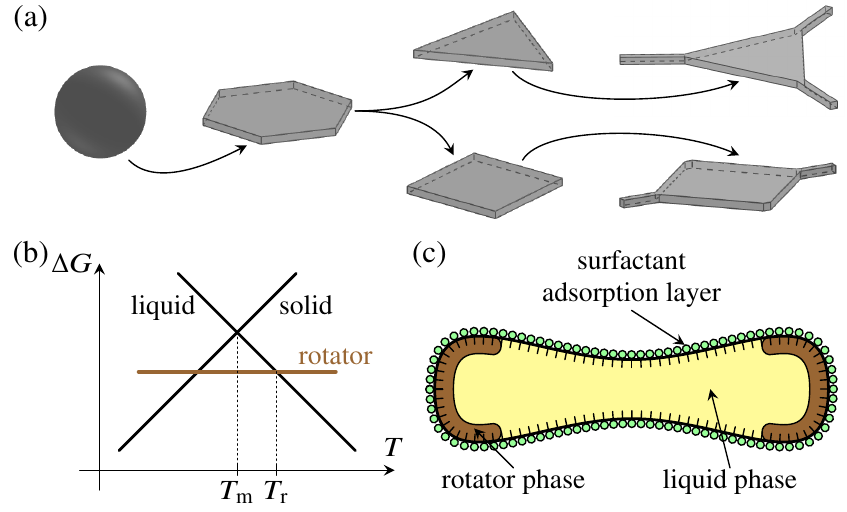}
\caption{Shape Shifting Droplets. (a)~Main stages of droplet shape evolution, following~\cite{denkov15}. Initially spherical 
droplets flatten, first into hexagons, then into quadrilaterals or triangles, before growing thin protrusions. (b)~Energy diagram, 
showing a range of temperatures close to the melting temperature where the formation of a rotator phase is energetically favored. 
(c)~Cross-section of flattened droplet. A layer of plastic rotator phase has formed at the edges of the droplet.} 
\label{fig1}
\end{figure}

Shape generating chemical systems are a very literal instantiation of Turing's quest for the chemical basis of morphogenesis \cite{turing}.   
Since they are to living systems what the shadows perceived by the eponymous cavemen of Plato's allegory \cite{plato} are to reality, it 
is the apparent simplicity of these systems that makes them ideal candidates for untangling the physical underpinnings of morphogenesis.  
A promising exemplar of such a system, revealed in a recent paper by Denkov \emph{et~al.} \cite{denkov15}, are small, micron-sized oil 
droplets suspended in an aqueous surfactant solution: as the temperature of the solution is slowly lowered towards the bulk freezing 
point of the oil, the initially spherical droplets undergo a remarkable array of shape transformations. The droplets flatten, and 
evolve through a sequence of polygonal shapes to first become hexagons, and later triangles or quadrilaterals (Fig.~\ref{fig1}a). 
At later stages, droplets grow thin protrusions, before ultimately thinning out into filaments.

This host of deformations was attributed to freezing of the surfactant layer before the bulk of the droplet, upon which the 
formation of a plastic rotator phase \cite{sirota} of self-assembled oil molecules next to the drop surface becomes energetically 
favorable, with long-range translational order (Fig.~\ref{fig1}b). This rotator phase rearranges into a frame of plastic rods at 
the drop perimeter which supports the polygonal structure (Fig.~\ref{fig1}c). Simple energy arguments~\cite{denkov15} were used 
to show that it is this rotator phase that must provide the energy to overcome the surface tension of the droplets and deform the latter. 
Other work~\cite{guttman16a} contends that these and similar deformations observed in related systems are the result of surface 
monolayer freezing only, however, and this was further discussed in a series of very recent papers \cite{guttman16b,cholakova16,denkov16}. 
The picture that emerges from this analysis is that the initial deformations of the spherical droplets to polyhedral shapes may be 
driven by the topological defects and the associated elastic scalings \cite{bowick,lidmar} resulting from the freezing of a planar, 
multi- or monomolecular layer, before the formation of tubules of rotator phase stretches and flattens the droplets. The geometric 
and energetic balances underlying the transitions between different polygonal stages and the later formation of protrusions, both 
driven by the expanding rotator phase, have so far remained unclear.

\begin{figure*}[t]
\includegraphics{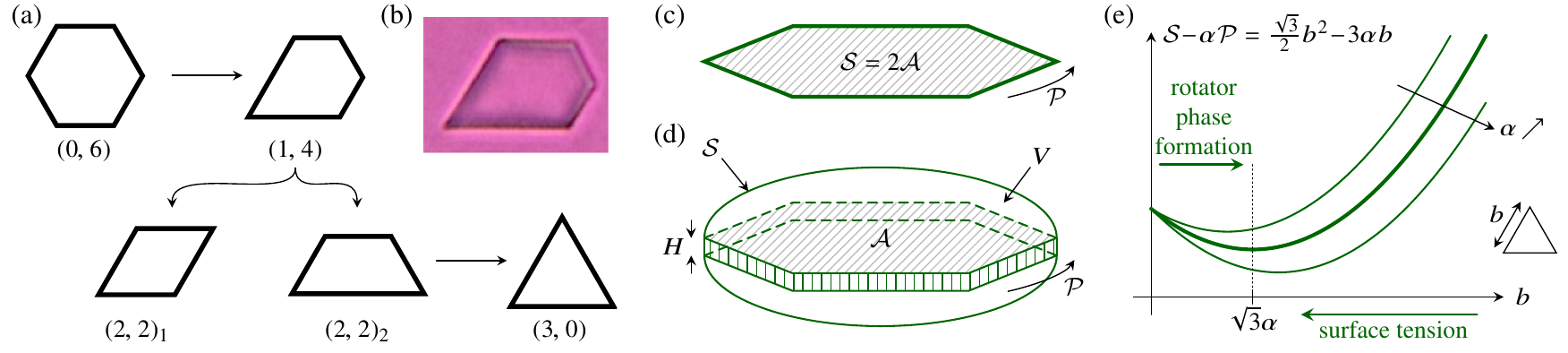}
\caption{Model of Shape Shifting Droplets. (a)~Classification of polygonal shapes with interior angles of $60^\circ$ or $120^\circ$, 
and transitions between shapes. (b)~Experimental image of pentagon state. (c)~Simplest droplet model: a polygonal frame of area 
$\mathcal{A}$ and perimeter $\mathcal{P}$ defines a flat droplet of surface area $\mathcal{S}=2\mathcal{A}$. (d)~Extended droplet 
model: conservation of the droplet volume $V$ and the finite height $H$ of the rim of rotator phase modify the relation 
between $\mathcal{S}$ and $\mathcal{A}$. (e)~Droplet energy: effect of competing surface tension and tendency to form rotator 
phase illustrated for the simplest, triangular droplet shape.}
\label{fig2}
\end{figure*}

Here, we analyze the competition between formation of the rotator phase and surface tension in the dynamics of these polygonal 
droplets. Our starting point is the experimental observation that the edges of the polygonal droplets are nearly straight, 
suggestive of a high energy cost to bend, and thus an even higher cost to stretch the rotator phase. We therefore assume 
the rotator phase to form straight, rigid rods at this point, but relax this assumption in later parts of the analysis. 
Vertices of the polygonal droplets correspond to defects in the rotator phase. A very large majority of the interior 
angles of the polygons are seen in experiments to have measures close to $60^\circ$ or $120^\circ$, suggesting that the 
defect energy landscape has strong minima near these values, which we therefore assume to be the only possible 
interior angles. 
Let $u,v$ denote, respectively, the number of vertices with angles of $60^\circ$ and $120^\circ$, 
so that if the polygons are convex,
\begin{align}
60^\circ\times u+120^\circ\times v=180^\circ\times(u+v-2).
\end{align}
The resulting equation, $2u+v=6$, has four solutions over the non-negative integers,
\begin{align}
(u,v)\quad =\quad (0,6),\;(1,4),\;(2,2),\;(3,0).
\end{align}
There are five polygonal shapes satisfying these requirements, shown in Figs.~\ref{fig2}a,b: equiangular hexagons, 
pentagons, two types of quadrilaterals (isosceles trapezoids and parallelograms) and equilateral triangles. These 
shapes have variable aspect ratios, and by contracting one of the sides to zero while extending the neighboring sides, 
hexagons can evolve into pentagons, and pentagons can evolve into either trapezoids or parallelograms. Of the latter two, 
only the trapezoids can evolve into triangles. This recovers the experimental sequence of shapes, and in particular the 
observed dichotomy between triangles and quadrilaterals as final outcomes (Fig.~\ref{fig1}a).

To study the evolution of the droplets, we turn to their energetics. Let $V$ be the 
droplet volume, assumed fixed, and $S$ its curved surface area, and assume that 
$V$ is the sum of the volumes of the rotator and liquid phases, 
$V=V_{\mathrm{r}}+V_{\mathrm{l}}$, and that their densities are equal. The energy of the droplet 
is 
\begin{align}
E=\gamma S+\mu_{\mathrm{l}}V_{\mathrm{l}}+\mu_{\mathrm{r}}V_{\mathrm{r}}+D. 
\end{align}
These are, respectively, (i) the surface energy with a coefficient $\gamma$ of surface tension \cite{surfacetensionnote}; 
(ii) the energies of both phases, proportional to their respective volumes, with chemical potentials $\mu_{\mathrm{r}}$, $\mu_{\mathrm{l}}$ 
per unit volume, where $\mu_{\mathrm{r}}<\mu_{\mathrm{l}}$ for the phase change to the rotator phase to be energetically favorable; 
(iii)~the defect energy $D$, which is a sum of contributions from each vertex. Only the defect energy is discrete; 
while it has constrained the above analysis of possible shapes, it does not affect the dynamics.

Next, we approximate $V_{\mathrm{r}}=A_{\mathrm{r}}P$, where $P$ is the perimeter of the 
polygonal frame, and $A_{\mathrm{r}}$ is a characteristic cross-sectional area of the rotator phase, depending 
on the intrinsic properties of the oil and the surfactant. Omitting constant terms and rescaling with the perimeter $P_0=(6\pi^2 V)^{1/3}$ 
of the initial spherical droplet we obtain the dimensionless energy $\mathcal{E}=E/\gamma P_0^2$, 
\begin{align}
\mathcal{E}=\mathcal{S}-\alpha \mathcal{P}, \qquad\alpha=\dfrac{A_{\mathrm{r}}}{V^{1/3}}
\dfrac{\mu_{\mathrm{l}}-\mu_{\mathrm{r}}}{(6\pi^2)^{1/3}\gamma}. \label{eq:E}
\end{align}
where $\mathcal{S}=S/P_0^2$
and $\mathcal{P}=P/P_0$.  
The dimensionless $\alpha>0$ is the key parameter of our analysis. For simplicity, we shall assume it to be constant; in 
the experimental system, it varies because of the variation of $\Delta\mu = \mu_{\mathrm{l}}-\mu_{\mathrm{r}}$ with 
temperature (and thus with time), the variations of $A_{\mathrm{r}}$ resulting from the kinetics of phase change, and 
finally, to a lesser extent, the weak variation of ${\gamma\approx 5\,\mathrm{mN/m}}$ with temperature \cite{cholakova16}.

\begin{figure*}[t]
\includegraphics{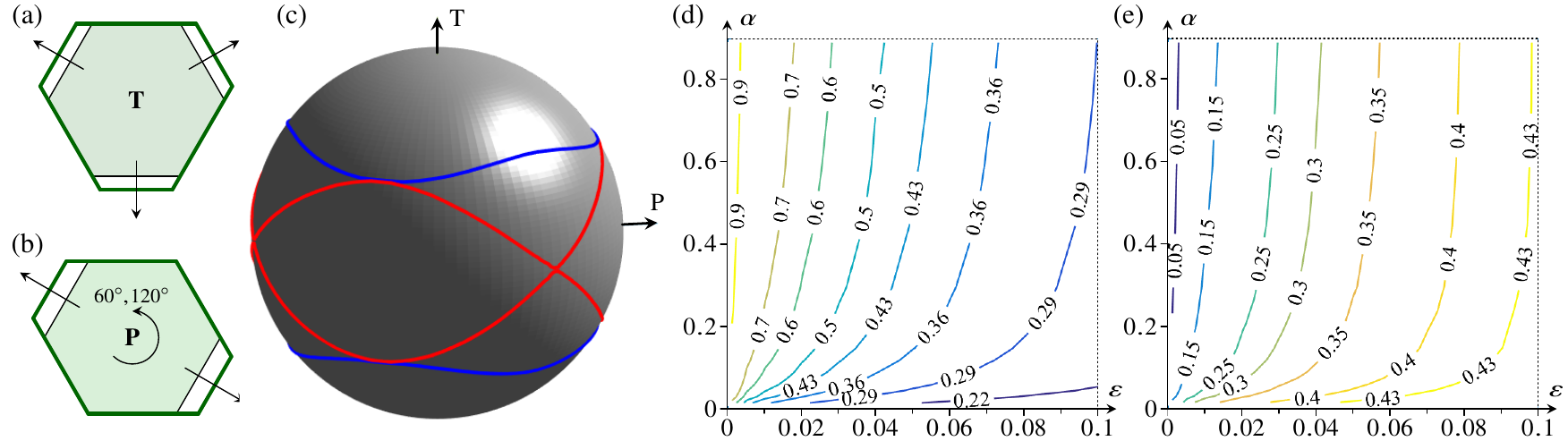}
\caption{Droplet Modes and Shape Statistics. (a)~Triangle mode (T), with eigenvalue $1$. (b)~Parallelogram mode (P), with eigenvalue~$2/3$. There are two linearly independent modes of this ilk; others are obtained by rotation by $60^\circ$ or $120^\circ$. (c)~Shape boundaries between triangles, trapezoids and parallelograms for perturbations of fixed magnitude $\varepsilon$ around the hexagon state, with eigendirections T and P indicated. Parameter values: $\alpha=0.3$, $\varepsilon=0.04$. (d)~Proportion of initial hexagons that evolve into triangles in $(\varepsilon,\alpha)$ space. (e)~Proportion of initial hexagons that evolve into parallelograms in $(\varepsilon,\alpha)$ space.} 
\label{fig3}
\end{figure*}

Approximating the cross section of the rotator phase as a semi-circular ring of outer radius $r\approx 1\,\text{\textmu m}$ and 
thickness $h\approx 100\,\mathrm{nm}$ \cite{denkov15,cholakova16}, we estimate $A_{\mathrm{r}}\approx0.3\,\text{\textmu m}^2$. 
At the temperature $T_{\mathrm{r}}$ of the liquid-to-rotator phase transition (Fig.~\ref{fig1}b), the two phases are in 
equilibrium, with $\Delta\mu=0$. When the system is cooled below $T_{\mathrm{r}}$,  
$\Delta\mu = \Delta S_{\mathrm{lr}}(T_{\mathrm{r}} - T)$, where 
$\Delta S_{\mathrm{lr}}\approx 6\cdot 10^5\,\mathrm{N/m^2K}$ is the entropy of transition per unit volume from the liquid to the 
rotator phase~\cite{kraack00} and $T_{\mathrm{r}} - T\approx 1\,\mathrm{K}$ during the platelet stage is sufficient to overcome 
the energy penalty incurred because of the expanding droplet surface. For initial droplet radii of $1-50\,\text{\textmu m}$ \cite{denkov15}, we 
thus estimate $\alpha\approx 5.7-0.1$.

The next step is to relate the surface area $\mathcal{S}$ to the geometry of the polygonal frame, which spans an 
area~$\mathcal{A}$. In the simplest approximation, we consider a flat two-dimensional droplet (Fig.~\ref{fig2}c), 
for which $\mathcal{S}=2\mathcal{A}$. This is the leading-order term of an expansion of $\mathcal{S}$  
with corrections \cite{SM} arising from volume conservation and the thickness $H\approx 2\,\text{\textmu m}$ \cite{denkov15} of the rim of 
rotator phase (Fig.~\ref{fig2}d) that we shall consider later.

The competition captured by (\ref{eq:E}) between the rotator phase and surface tension is illustrated for 
equilateral triangles in Fig.~\ref{fig2}e, where $\mathcal{E}=\sqrt{3}b^2/2-3\alpha b$, with $b$ the edge length.
Surface tension acts to decrease the side lengths of the triangle, which the formation of the rotator phase tends to expand. 
For the other shapes, disproportionation of edge lengths gives additional degrees of freedom, and we must derive the equations 
governing the shape dynamics to explore this aspect. We impose a standard constitutive law~\cite{suo97}, that the normal velocity 
of the perimeter is proportional to the thermodynamic pressure resulting from the corresponding energy variations, and derive 
the corresponding equations in the Supplemental Material \cite{SM}. Videos showing the numerically computed evolution of hexagons 
into triangles, parallelograms or trapezoids are available online \cite{SM}; whether a trapezoid evolves into a triangle or 
not depends on the aspect ratio of the droplet when it reaches the trapezoid state.

A linear stability analysis around the regular hexagonal shape \cite{SM} reveals three linearly independent growing modes: 
a triangle mode (Fig.~\ref{fig3}a; scaled eigenvalue $1$) and two parallelogram modes (Fig.~\ref{fig3}b; eigenvalue $2/3$). We 
take these eigenmodes as a basis for perturbations of prescribed magnitude $\varepsilon$ around the regular hexagon, and 
map the perturbations yielding, respectively triangles, parallelograms or trapezoids onto the surface of a sphere~(Fig.~\ref{fig3}c). 
Varying the parameters $\alpha$ (in the experimental range determined above) and $\varepsilon$, we compute the respective extents of 
these regions and thus the proportion of initial hexagons evolving into triangles and parallelograms (Fig.~\ref{fig3}d,e): 
at small~$\varepsilon$, hexagons are overwhelmingly likely to evolve into triangles, but as $\varepsilon$ is increased, parallelograms and 
trapezoids dominate, with trapezoids somewhat less likely to arise than parallelograms. To relate this to experiments, we observe 
that $\varepsilon$ should increase with increasing cooling rate: at faster cooling, the arrangement of topological defects on the surface 
of the spherical droplet is less regular at the time when the rotator phase starts to form. Thus the higher the cooling rate, the less 
likely droplets (of the same initial size) are to evolve into triangles. Similarly, at the same cooling rate, given 
that $\alpha\propto V^{-1/3}$, larger droplets are less likely to evolve into triangles. This is borne out qualitatively by the experiments
\cite{denkov15,cholakova16}.

\begin{figure}[t]
\includegraphics{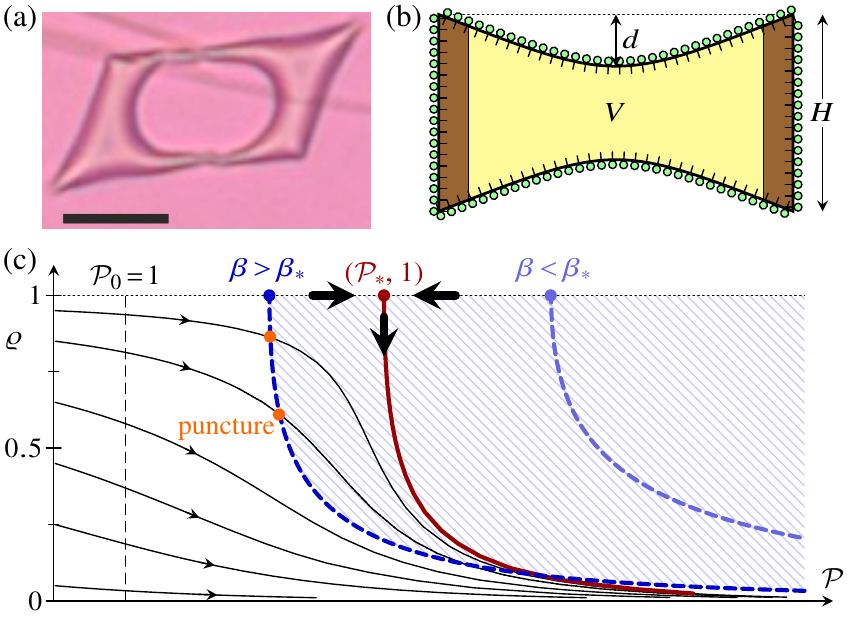}
\caption{Droplet Puncture. (a)~Punctured parallelogram as outcome of droplet shape evolution. Scale bar: $20\,\text{\textmu m}$. 
(b)~Idealised cross-section of an inverted droplet. (c)~Trajectories of (\ref{eq:dynsys}) indicated by solid lines, overlaid 
with the puncture condition (\ref{eq:puncture}), represented by dashed lines. Thick solid line: trajectory through saddle. 
Parameter values: $\alpha=0.3$, $H/V^{1/3}=0.25,0.047$.}
\label{fig4}
\end{figure}

We now turn to the topological transition of droplet puncture: experimentally, some parallelograms elongate into thin rods, 
while others puncture in their center before freezing (Fig.~\ref{fig4}a), a strong experimental signature of the formation of 
rotator phase next to the perimeter~\cite{denkov15}. A parallelogram with side lengths $a,b$ is most conveniently described in terms of 
its perimeter $\mathcal{P}=2(a+b)$ and aspect ratio $\varrho=a/b$. By symmetry, we may restrict the analysis to $0<\varrho\leqslant 1$. Droplet 
puncture occurs as the droplet surface inverts because of the finite thickness~$H$ of the rim of rotator phase (Fig.~\ref{fig4}b), and 
when, at a given aspect ratio, the depth $d$ of the inverted cap reaches the value $H/2$, or equivalently, when 
the droplet exceeds the critical perimeter~\cite{SM}
\begin{align}
\mathcal{P}_{\mathrm{crit}}(\varrho)=C\dfrac{V^{1/6}}{H^{1/2}}\dfrac{1+\varrho}{\sqrt{\varrho}}g(\varrho), \label{eq:puncture}
\end{align}
where $C\approx 0.75$ is a numerical constant, and where $g(\varrho)$ is a decreasing function of $\varrho$, with $g(1)=1$, that 
must be evaluated numerically, although results do not change qualitatively if it is taken to be constant \cite{SM}. The evolution 
of $\mathcal{P}$ and $\varrho$ is governed by~\cite{SM}
\begin{align}
\dot{\mathcal{P}}&=8\sqrt{3}\left\{\dfrac{2\alpha(1+\varrho)^2}{\sqrt{3}\mathcal{P}\varrho}-1\right\},&&\dot{\varrho}=-\dfrac{4\sqrt{3}}{\mathcal{P}}\left(1-\varrho^2\right),\label{eq:dynsys}
\end{align}
which has a fixed point at $(\mathcal{P},\varrho)=(\mathcal{P}_\ast,1)$, with ${\mathcal{P}_\ast=8\alpha/\sqrt{3}}$. Fig.~\ref{fig4}c shows a 
partial phase portrait of this dynamical system, overlaid with the puncture condition (\ref{eq:puncture}). If 
$\mathcal{P}_{\mathrm{crit}}(1)>\mathcal{P}_\ast$, droplets cannot puncture, while, if $\mathcal{P}_{\mathrm{crit}}(1)<\mathcal{P}_\ast$, 
only droplets with small $\varrho$ will not puncture (Fig.~\ref{fig4}c). The condition $\mathcal{P}_{\mathrm{crit}}(1)>\mathcal{P}_\ast$ 
rearranges to $\beta\equiv\alpha H^{1/2}/V^{1/6}<\beta_\ast\approx 0.33$ in terms of the physical parameters. Since $\beta\propto V^{-1/2}$, 
this means that larger droplets will puncture more rarely. Experiments, however, show that the smallest droplets very rapidly elongate into 
rods without puncture, hinting that
further understanding of the phase change kinetics must be gained to explain the size dependence of this transition.

\begin{figure}[t]
\includegraphics{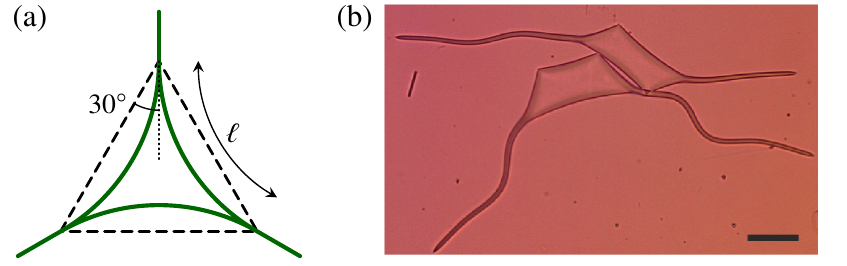}
\caption{Growth of Protrusions. (a)~Geometry: bending of sides removes defects and allows protrusions to grow. 
(b)~Parallelograms and trapezoids sprout protrusions from $60^\circ$ angles, but not from $120^\circ$ angles. Scale bar: $20$ \textmu m.}
\label{fig5}
\end{figure}

Finally, we address the formation of protrusions sprouting from the vertices of the polygons. To this end, 
we must relax our initial assumption that the droplets have straight rigid sides, for otherwise the formation of a protrusion 
corresponds to replacing one defect with two defects (one of $120^\circ$ with two $120^\circ$ ones, or even one of $60^\circ$ 
with two $150^\circ$ ones), which is energetically unfavorable \cite{protnote}. However, by the time protrusions appear 
in the experimental system, the sides of the polygons are bent inwards slightly. If we therefore allow the sides to bend elastically, 
the defects at the vertices may be removed by bending (Fig.~\ref{fig5}a), allowing the growth of protrusions.

The energetics are particularly simple for triangles: bending each of the sides of an equilateral triangle of 
sidelength $\ell$ into a circular segment of length $\ell$ intercepting an angle of $60^\circ$ on a circle of radius $3\ell/\pi$ 
removes the defects at the corners (Fig.~\ref{fig5}a), and reduces the surface area by an amount $\sim\ell^2$ at a bending energy 
cost scaling as $\ell(1/\ell)^2\sim 1/\ell$. Hence, if $\ell$ (or, equivalently, $\mathcal{P}$) is large enough, 
the disappearance of the defects is energetically favorable. The same criterion applies, perhaps counterintuitively, to 
parallelograms or trapezoids~\cite{SM}. It is however clear that more bending is required to remove a defect of $120^\circ$ than 
one of $60^\circ$, and hence we expect protrusions to grow only from the acute angles of parallelograms or trapezoids, as 
observed (Fig.~\ref{fig5}b).

In the simplest model, protrusions grow either indefinitely, or do not grow at all: consider a parallelogram with two equal 
protrusions, each of length $p$ and let $\ell_{\mathrm{p}}$ denote the cross-sectional extent of the protrusions. In line with our previous 
assumptions on the rotator phase, we assume $\ell_{\mathrm{p}}$ to depend on the material properties only, and not on the shape. 
The energy is thus
\begin{align}
\mathcal{E}=\mathcal{E}_0+4(\ell_{\mathrm{p}}-\alpha)p, 
\end{align}
where $\mathcal{E}_0=2\mathcal{A}-\alpha\mathcal{P}$ is the energy in the absence of protrusions. In particular, 
if $\alpha>\ell_{\mathrm{p}}$, protrusions grow \emph{ad infinitum} and do not grow at all otherwise. Experiments reveal, however, 
that protrusions in parallelograms may shrink and disappear again as the parallelogram elongates into a thin rod. 
This effect results from volume conservation and the three-dimensional structure of the droplet (Fig.~\ref{fig4}b): after the 
droplet has inverted, shrinking the protrusions reduces the surface area (both of the protrusions and of the bulk of the droplet, 
because volume conservation means that its surface curves inwards less), but does incur a cost of reducing $\mathcal{P}$. 
This competition may cause the protrusions to shrink; in particular, the surface area reduction is larger the 
smaller is $\mathcal{A}$ (which is the reason why this effect is observed in elongated parallelograms rather than triangles, for instance). 
This effect is further enhanced, at the same $\mathcal{A}$, for more elongated parallelograms~\cite{SM}.

In this Letter, we have shown how the simplest competition between a single phase transition and surface tension in a discrete geometry 
affords a wealth of shape transitions and reproduces and explains many of the experimental findings \cite{denkov15,cholakova16}. 
Taking the analysis from the plane to three dimensions and understanding the geometric basis for the flattening of the droplets 
is a key challenge for future work, as are the kinetics of the phase change driving the shape evolution.

\begin{acknowledgments}
We thank S.~Tcholakova and I.~Lesov for discussions and experimental data. PAH and SKS are grateful to the Department of 
Chemical and Pharmaceutical Engineering of the University of Sofia for their hospitality during the annual retreat of 
the department at Giolechitsa (Bulgaria) in March 2016. This work was supported in part by the Engineering and Physical Sciences Research
Council (PAH), an Established Career Fellowship from the EPSRC (REG), and the European Research Council (grant EMATTER \#280078 to SKS).
\end{acknowledgments}


\begin{thebibliography}{10}
\bibitem{surfacetensionnote}
This assumes that the coefficients of surface tension between the water on the one hand and the liquid or rotator
phases on the other are equal, and ignores the surface tension between the two oil phases (or alternatively that the
interfacial tension water-liquid phase is approximately equal to the sum of the interfacial tensions between the water and rotator phases and
the rotator and liquid phases). These additional effects would but modify slightly the final parameter, as would taking into account the density difference between the two phases.
\bibitem{protnote}
This assumes symmetric defects; one could however conceive of protrusions that are aligned with one of the edges at the vertex whence they 
sprout. Such protrusions are observed but rarely, suggesting that these `composite' defects are energetically unfavorable, too.
\end{thebibliography}
\end{document}